\documentclass[aps,prl,twocolumn,amsfonts,amssymb,amsmath,floatfix,showpacs]{revtex4}
\usepackage{graphics}
\usepackage{color}

\newcommand{\eq}[1]{(\ref{#1})}         
\newcommand{\etal}[1]
  {{\it et al.\/}\ifx#1.\else\expandafter#1\fi}
\newcommand{\Fig}[1]{Fig.~\ref{#1}}     
\newcommand{\fig}[1]{fig.~\ref{#1}}     
\newcommand{\ie}[1]
  {{\it i.e.\/}\ifx#1.\else\expandafter#1\fi}
\newcommand{\dblfigure}[3]
  {\begin{figure*}[tbp]#1\caption[]{#2}\label{#3}\end{figure*}}
\newcommand{\sglfigure}[3]
  {\begin{figure}[tbp]#1\caption[]{#2}\label{#3}\end{figure}}

\renewcommand{\@}{\partial}             
\newcommand{\const}{\mathrm{const}}     
\renewcommand\d{{\mathrm d}}            
\newcommand{\Df}[2]{\frac{\d{#1}}{\d{#2}}} 
\newcommand{\e}{\mathrm{e}}             
\newcommand{\Heav}{H}                   
\renewcommand{\Im}[1]{{\rm Im}\left(#1\right)}	
\renewcommand{\i}{\mathrm{i}}           
\newcommand{\Mx}[1]{
\left[\begin{array}{cccccccc}#1\end{array}\right]}
\newcommand{\mx}[1]{\mathbf{#1}}        
\renewcommand{\Re}[1]{{\rm Re}\left(#1\right)}	
\newcommand{\Real}{\mathbb{R}}          
\newcommand{\T}{^{\mathrm{T}}}          

\newcommand{\D}{\mx{D}}                 
\newcommand{\db}{\delta_\parb}          
\newcommand{\de}{\delta_\pareps}        
\newcommand{\dist}{d}                   
\newcommand{\evec}{\hat{\mx{e}}}              
\newcommand{\F}{F}                      
\newcommand{\f}{\mx{f}}                 
\newcommand{\h}{\mx{h}}                 
\renewcommand{\L}{\mathcal{L}}          
\newcommand{\Lp}{\L^{+}}                
\newcommand{\Omg}{\Omega}               
\renewcommand{\P}{P}                    
\newcommand{\p}{\mx{p}}                 
\newcommand{\para}{a}                   
\newcommand{\parb}{b}                   
\newcommand{\pareps}{\epsilon}          
\newcommand{\R}{{\vec R}}               
\newcommand{\Ri}{R_i}                   
\newcommand{\RF}{\mx{\W}}               
\renewcommand{\r}{{\vec r}}             
\newcommand{\strength}{\beta}           
\newcommand{\U}{\mx{U}}                 
\renewcommand{\u}{\mx{u}}               
\renewcommand{\vr}{\F_r}                
\newcommand{\vt}{\F_a}                  
\newcommand{\W}{W}                      
\newcommand{\X}{X}                      
\newcommand{\Y}{Y}                      
\newcommand{\Z}{R}                      
\newcommand{\xs}{x_*}                   
\newcommand{\ys}{y_*}                   

\newcommand{\dt}{\Delta t}              
\newcommand{\dx}{\Delta x}              

\begin{document}

\title{Orbital motion of spiral waves in excitable media}

\author{V.~N.~Biktashev}
\affiliation{Department of Mathematical Sciences, University of Liverpool, Liverpool L69 7ZL, UK}

\author{D.~Barkley}
\affiliation{Mathematics Institute, University of Warwick, Coventry CV4 7AL, UK}

\author{I.~V.~Biktasheva}
\affiliation{Department of Computer Science, University of Liverpool, Liverpool L69 3BX, UK}

\date{\today}
\begin{abstract}
  Spiral waves in active media react to small perturbations as
  particle-like objects. Here we apply the asymptotic theory to
  the interaction of spiral waves with a localized inhomogeneity,
  which leads to
  a novel prediction: drift of the spiral rotation centre
  along circular orbits around the inhomogeneity. The stationary
  orbits have alternating stability and fixed radii, 
  determined by the properties of the bulk medium and the type
  of inhomogeneity, while the drift speed along an orbit depends on
  the strength of the inhomogeneity.  Direct simulations confirm the
  validity and robustness of the theoretical predictions and show that
  these unexpected effects should be observable in experiment.
\end{abstract}
\maketitle

The interest in the dynamics of spiral waves as regimes of
self-organization has considerably broadened in
the last decades, as they have been found in ever more physical
systems of diverse types (%
magnetic films~\cite{Shagalov-1997}, %
liquid crystals~\cite{Oswald-Dequidt-2008}, %
nonlinear optics~\cite{LeBerre-etal-2005}, %
new chemical systems~\cite{Agladze-Steinbock-2000}, %
and in 
population~\cite{Igoshin-etal-2004}, %
tissue~\cite{Dahlem-Mueller-2003}, and %
subcellular~\cite{Bretschneider-etal-2009} 
biology).
In a perfectly uniform medium the core of a spiral wave may be
anywhere, depending on initial conditions. However, real systems are
always heterogeneous, and therefore spiral drift due to inhomogeneity
is of great practical interest to applications.  Understandably, such
drift has been mostly studied in excitable chemical reactions and the
heart, where drift due to a gradient of medium properties
\cite{%
  Fast-Pertsov-1992,
  Luengviriya-etal-2006
} and pinning \endnote{
  Note that the term ``pinning'' can also be used in a completely difference sense,
  see~\cite{Jensen-etal-1994}, although ``self-pinning''
  perhaps would be more accurate in that case. 
} (anchoring, trapping) to a localized inhomogeneity
\cite{%
  Nettesheim-etal-1993,
  Pertsov-etal-1993,
  Lim-etal-2006
} have been observed in experiments and simulations. Interaction with
localized inhomogeneity can be considered to be a
particular case of the general phenomenon of vortex pinning to material
defects~\cite{Lugomer-etal-2007}.

Here we identify a new type of spiral wave dynamics: precession around
a localized inhomogeneity along a stable circular orbit.  We predict
this novel
phenomenon theoretically, describe its key features,
and confirm it by numerical simulations. We argue that this orbital
movement of spiral waves is robust and prevalent, has nontrivial
and surprising consequences for applications and
should be directly observable in experiments.

We consider reaction-diffusion equations, which is the most popular
class of models describing spiral waves:
\begin{equation}
\@_t\u = \f(\u,\p) + \D \nabla^2 \u, \quad 
     \label{RDS}
\end{equation}
where $\u,\f\in\Real^{\ell}$, $\D\in\Real^{\ell\times\ell}$, $\p\in\Real^m$,
$\u(\r,t)$ is the dynamic vector field, $\r\in\Real^2$,
$\p(\r)=\p_0+\p_1(\r)$, $|\p_1|\ll1$, is the vector of parameters,
$\D$ is diffusion matrix.
For $\p=\p_0=\const$, system \eq{RDS} is assumed to have 
spiral wave solutions rotating with angular velocity
$\omega$ (taken here to be clockwise for $\omega>0$),
\begin{equation}
\u = \U(\rho,\vartheta + \omega t - \Phi) ,
                                      \label{SW}
\end{equation}
where $(\rho,\vartheta)$ are polar coordinates defined with
respect to the center of rotation $\R=(\X,\Y)\T$, and $\Phi$
is the initial rotation phase.  

In the presense of a small perturbation $\p_1(\r) \ne 0$, the spiral's
center of rotation $\Z=\X+\i\Y$ is not
constant but slowly evolves with the
equation of motion
\begin{equation}
\Df{\Z}{t} = 
  \frac{\omega\e^{ \i \Phi }}{2\pi}
  \int\limits_{t-\pi/\omega}^{t+\pi/\omega} 
  \e^{-\i \omega\tau} 
 \iint\limits_{\Real^2}
    \left[\RF\left(\rho,\theta\right)\right]^{+}
 \h(\r,\tau) \,
  \d^2\r\;\d\tau,
                                        \label{forces}
\end{equation}
where $\rho=\rho(\r-\R)$ and $\theta=\vartheta
(\r-\R)+\omega\tau-\Phi$ are polar coordinates in the corotating frame
of reference, and $\h$ is the perturbation to the
  right-hand side of Eq.~\eq{RDS}.  Function $\RF$ is called
  the response function (RF) and defines the sensitivity of the spiral
  wave position with respect to perturbations in different
  places. Technically, $\RF$ is a projector onto the eigenmode
  corresponding to the neutral stability with respect to spatial
  translations and is calculated as the eigenfunction %
$\Lp \RF = -\i\omega \RF$
of the adjoint linearized operator
$\Lp=\D\T\nabla^2+\omega\@_\theta+\left(\@_{\u}\f(\U;\p_0)\right)\T$, 
see for details 
\cite{Biktashev-Holden-1995,Biktasheva-Biktashev-2003,Biktasheva-etal-2010}.

\dblfigure{\includegraphics{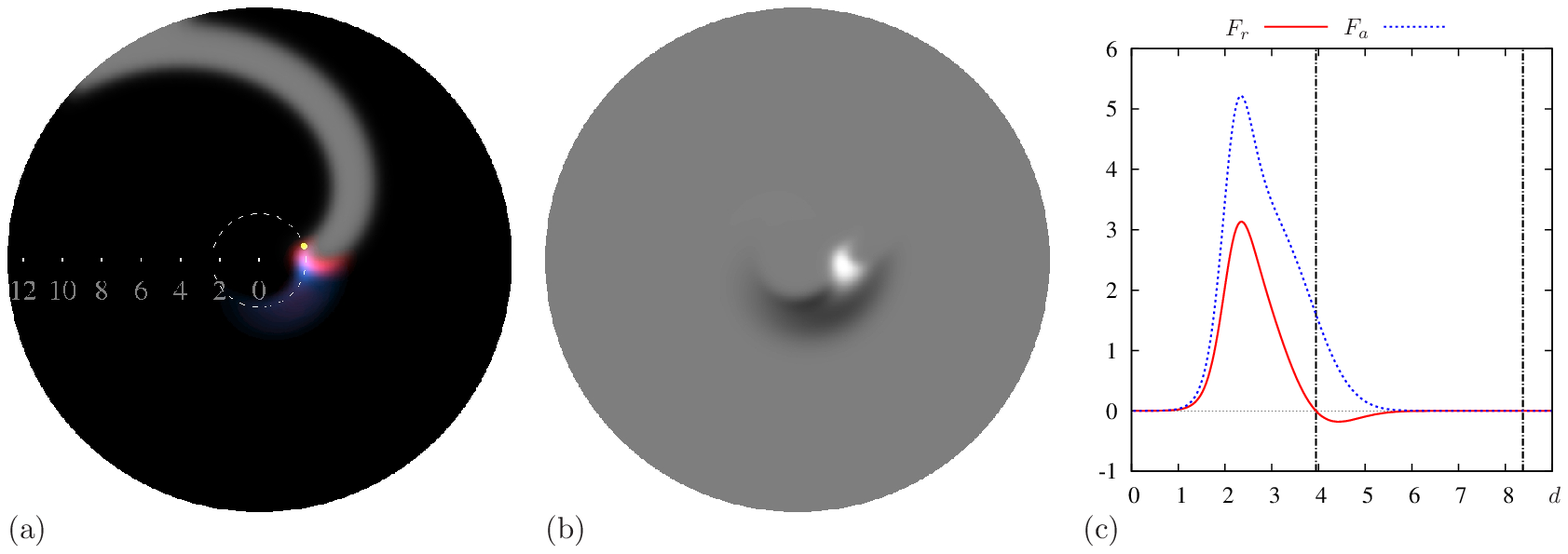}}{
   (color online) 
  (a) Response function for a spiral wave.  The spiral is visualized with a
  gray-scale plot of the $u$ field. The yellow point indicates the spiral tip
  and the dashed white line shows its trajectory as the spiral
  rotates. Superimposed is the response function in term of the $u$
  component $|\W_u|$ (red) and the $v$ component $|\W_v|$ (blue).
  (b) Enhanced visualization of one component of the response
  function. $\Re{W_v}$ is plotted with medium gray zero (periphery), light
  gray positive and dark gray negative.
  (c) Drift force. The radial $\vr$ (solid red) and azimuthal $\vt$ (dashed
  blue) components of the drift force, as functions of the distance $\dist$,
  calculated by~\eq{central-force}.  The vertical dash-dotted lines show zeros
  of the radial component, corresponding to the radii at which stationary
  orbital movement of the spiral wave is possible.

}{theory}

For an inhomogeneity
$\h = \@_{\p}\f(\U(\rho,\theta);\p_0) \, \p_1(\r)$ uniform inside a
  disk of radius $\Ri$, 
  $\p_1(\r)=\frac{\strength}{\pi\Ri^2}\,\Heav(\Ri-|\r|)\,\evec$,
  $\strength\ll1$, where
  $\Heav()$ is the Heaviside step function and 
  $\evec\in\Real^m$, $||\evec||=1$,
  equation \eq{forces} gives
\begin{equation}
  \Df{\Z}{t} = - \strength \frac{\Z}{|\Z|} \F(|\Z|) . \label{EoM}
\end{equation}
Here $F$ is the ``drift force'', defined as the
drift velocity per unit
inhomogeneity strength $\strength$. 
For small $\Ri$, the expression for $F$ simplifies to
\begin{equation}
  \F(\dist) = \int\limits_0^{2\pi} \e^{-\i\theta}\,
  \left[\RF(\dist,\theta)\right]^+ \, \@_{\p}\f(\U(\dist,\theta);\p_0) \, \evec
  \,\frac{\d\theta}{2\pi}.
                                        \label{central-force}
\end{equation}

We calculated the spiral wave solution $\U$ and the response function
$\RF$ using the method described in
\cite{Biktasheva-etal-2009,epaps} for the
Barkley~\cite{Barkley-1991} kinetics
$\u=(u,v)$, $\p=(\para,\parb,\pareps)$, 
$\f=(f_u,f_v)\T$,
$f_u=\pareps^{-1}u(1-u)(u-(v+\parb)/\para)$, $f_v=u-v$,
$\para_0=0.7$,
$\parb_0=0.1$,
$\pareps_0=0.02$,
and $\D=\Mx{1&0\\0&0}$.
This model is ``excitable'', that is, it has a unique
  spatially uniform steady state, stable with respect to small
  perturbations~\cite{Kness-etal-1992}. 
At the chosen parameter values, the spiral wave
  solutions are stable~\cite{Barkley-1994}.
 \Fig{theory}(a,b) shows $\W_u$ and $\W_v$ components and their
location relative to the spiral.
\Fig{theory}(c) shows graphs of the radial,
$\vr(\dist)=\Re{\F(\dist)}$ (positive for attraction) and azimuthal,
$\vt(\dist)=\Im{\F(\dist)}$ components of the drift force,
for the localized inhomogeneity in
parameter $\parb$, \ie\ $\evec=(0,1,0)\T$.

The essence of our new finding is that there is the change of sign of
radial force $\vr(\dist)$ at $\dist=\dist_1\approx3.95$. This
follows  from the sign changes of $\RF$ components as seen in
\fig{theory}(b).  For positive $\strength$, this means attraction to
inhomogeneity at small distances and repulsion at larger distances.
For negative $\strength$, however, there will be a repulsion from the
inhomogeneity at
$\dist<\dist_1$ and attraction at $\dist>\dist_1$,  so that
$\dist=\dist_1$ is a stable distance. The latter corresponds to the  
drift along an orbit of radius $\dist_1$ with the speed
$|\strength\vt(\dist_1)|$. There is a further root
of $\vr$ at $\dist=\dist_2\approx8.38$; however, the corresponding value
of $\vt$ is very small, $\sim10^{-9}$, so no drift is easily
observable there.

\dblfigure{\includegraphics{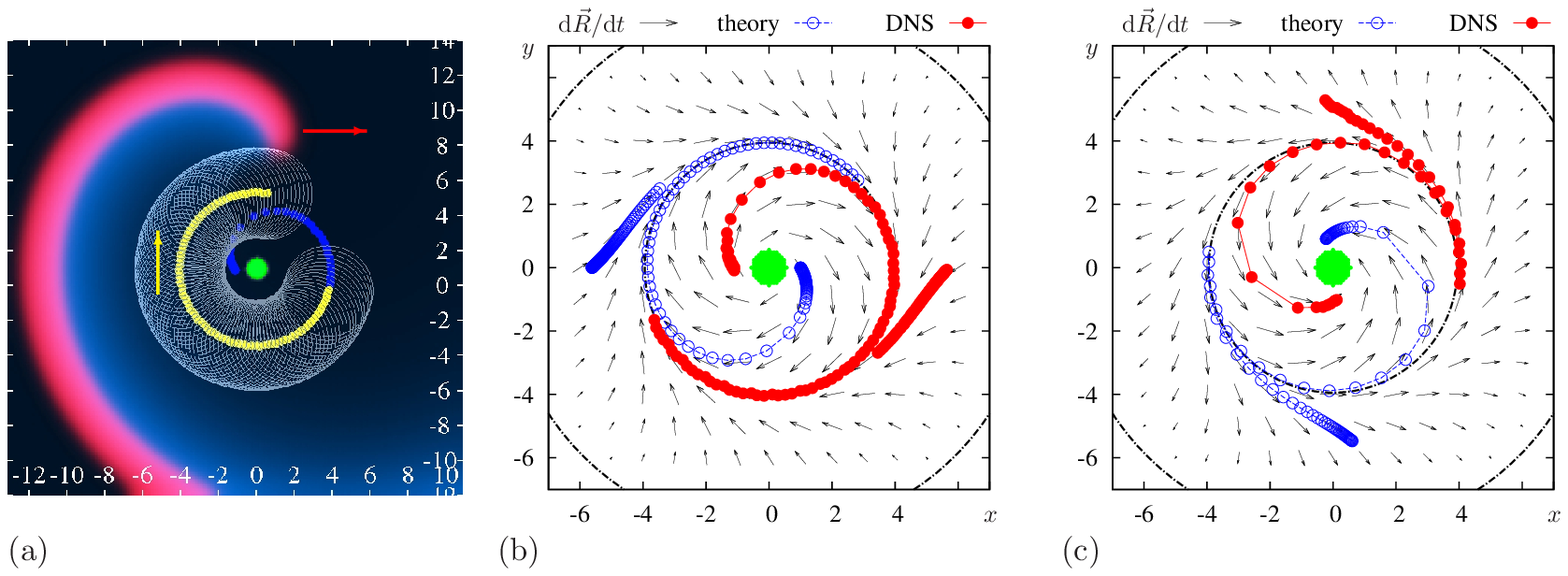}}{
(color online) Orbital movement of a spiral.
  (a) Simulation of a spiral wave in the presence of an inhomogeneity. The
  spiral starts near the inhomogeneity ($\db=-0.02$, green disk), is repelled
  from it, and launches into a stable circular clockwise orbit~\cite{epaps}.
  The $u$ field (red) and $v$ field (blue) of the final spiral are shown.  The
  preceding tip trajectory along the stable orbit is shown by a white
  line. The preceding centers rotation $\R$ are indicated both for this orbit
  (yellow) and the preceding evolution away from the inhomogeneity (blue).
  (b) Theoretical vector field (black arrows, nonlinearly scaled for
  visualization) and predicted trajectories (blue open circles) for the center
  of a spiral wave near an inhomogeneity (green disk). Actual trajectories for
  the spiral center from a DNS (red filled circles), with $\db=-0.001$.  Black
  dash-dotted circles indicate stationary orbits as predicted by theory.  Only
  every 20th position of the center is shown on both theoretical and DNS
  trajectories.
  (c) Same for $\db=0.003$. 

}{b-inhom}

\Fig{b-inhom} shows confirmation of the theoretical prediction of the
orbital movement by direct numerical simulations (DNS)~\cite{epaps}.
Panel (a) illustrates the relationship between the DNS spiral wave
solution, its tip and its instantaneous rotation centre, and panels (b)
and (c) show the centre trajectories predicted by the theory and
calculated by DNS, for  
$\Ri=0.56$ and different values of $\db=\strength/(\pi\Ri^2)$. 
Trajectories show circular orbits, 
attracting for $\db<0$ and repelling for $\db>0$, with the
radius 
 indistinguishable from $\dist_1$.

Panels (b) and (c) illustrate two key features of orbital drift: the
orbiting speed depends on the strength of the inhomogeneity, while the
radius of the orbit does not -- it depends only on the properties of
the unperturbed medium. These features follow from the theory and are
confirmed by DNS: trajectories in panel (c) have the same shape as in
panel (b), only the spirals drift along those trajectories faster
and in the opposite direction.

\sglfigure{\includegraphics{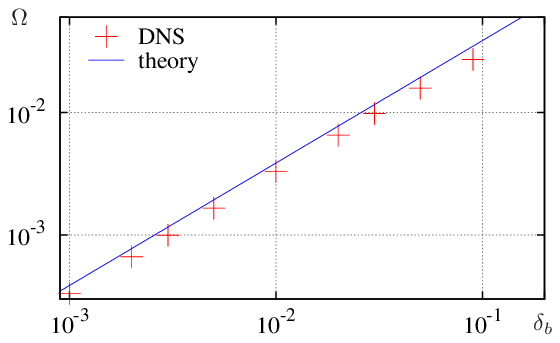}}{
  (color online) 
  Angular speed $\Omg$ of orbital movement of the spiral wave as
  a function of the amplitude of the parametric inhomogeneity $\db$:
  theoretical prediction vs
  measurements from direct numerical simulations. 
}{db-o}

\Fig{db-o} provides  quantitative comparison between theory and DNS.
The theoretical value of the angular velocity of the orbital movement
is $\Omg=\left|\db\pi\Ri^2 \vt(\dist_1)/\dist_1\right|$, implying
that $\Omg$ should vary linearly with $|\db|$ with slope
$\left|\pi\Ri^2 \vt(\dist_1)/\dist_1\right|$.
 Linearity of $\Omg(\db)$ is indeed found in the DNS,
remarkably up until $|\db/\parb_0|=0.9$. The ratio $\Omg/\db$ in simulations
is slightly smaller than the theoretical value, 
due to dicretization and approximations used. 

\dblfigure{\includegraphics{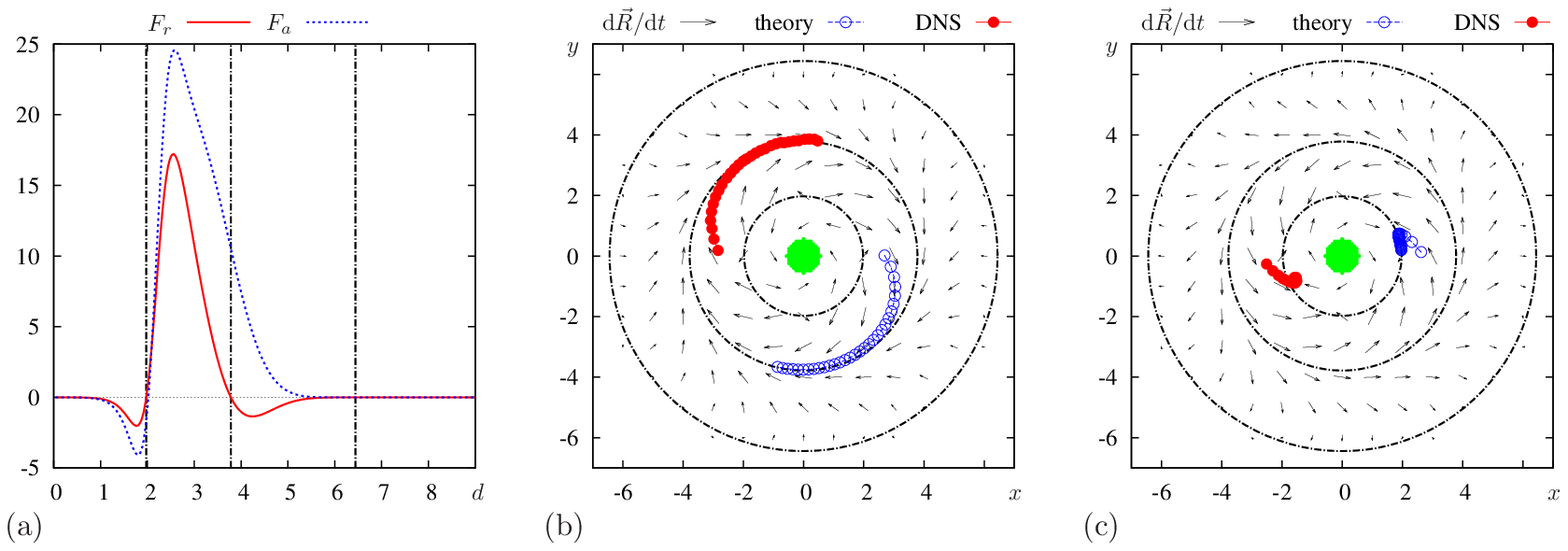}}{
  (color online) 
  Orbital movement of the spiral due to perturbation in parameter
  $\pareps$. 
  (a) Drift force components as functions of distance for this
  inhomogeneity. The notation is the same as in \fig{theory}(c).
  (b,c) Comparison of theoretical predictions and simulations, for
  (b) $\de=-0.001$ and (c) $\de=0.001$. 
  The notation is the same as in \fig{b-inhom}(b,c). 
  Shown are pieces of trajectories of the same temporal length $t=0\dots500$. 
}{e-inhom}

\Fig{e-inhom} compares theory and DNS for the perturbation in the
parameter $\pareps$ rather than $\parb$.  There are now three roots of
$\vr(\dist)$, namely $\dist_1\approx 1.97$, $\dist_2\approx 3.78$ and
$\dist_3\approx 6.45$, two of which are in the experimentally observable range.
Roots $\dist_1$ and $\dist_2$
have alternative stability: the $\dist_1$-orbit is unstable for
positive $\de$ and stable for negative $\de$, and $\dist_2$-orbit is
the other way round.  Thus a trajectory starting in between will enter
into an orbital motion in any case: to the outer orbit for $\de<0$ and
to the inner orbit for $\de>0$,  see
\fig{e-inhom}(b) and (c). 

In \fig{e-inhom}(a), root $\dist_1$ is very close to a root of
$\vt(\dist)$, and $\vt(\dist_1)$ is very small. Thus the inner orbit
is attracting for $\de>0$ but the movement along it is very slow
(\fig{e-inhom}(c)) and with short observation time, it may look as if
the spiral is attracted to any point at distance $\dist_1$ from the
inhomogeneity and stands still there.

To conclude, we have reported a new type of interaction of spiral
waves with localized inhomogeneities: orbital movement. 
This new interaction has key features that should be observable in
experiments, namely the orbiting speed depends on the strength of the
inhomogeneity, while the stationary orbit radii form a discrete set depending
only on the properties of the unperturbed medium. 
The phenomenon is rather generic: we have
found it in a number of other models and for more general shape of inhomogeneity 
\cite{Biktasheva-etal-2010}.
The possibility of orbital drift, related to a change of sign of an
equivalent of $\vr(\dist)$ was discussed at a speculative
level in~\cite{LeBlanc-Wulff-2000}; how often this phenomenon 
may occur in reality is a more complicated question. 
The equivalent of response functions calculated
in~\cite{Elkin-Biktashev-1999} has a structure which suggests that for
large-core spirals there is an infinite set of orbits.
In practice, orbital motion can only be observed for lower orbits where
the orbiting speed is noticeable. The orbits have alternating
stability, depending on the sign of the inhomogeneity. From this
viewpoint, ``pinning'' as considered in~\cite{Pazo-etal-2004} in
the same model as here, appears as a degenerate case of orbital
motion, with a zero radius.

In certain circumstances, while an orbiting spiral may have the same
macroscopic signature as a meandering one\endnote{%
  Orbiting may in fact have been observed by Zou \etal\
  \cite[p.R802]{Zou-etal-1993}, however it is difficult to be certain
  as no details were given. %
}, the microscopic details
leading to this motion are different.
Meander, in the proper sense, is
due to internal instabilities of a spiral wave, whereas orbital motion
is due to inhomogeneity. E.g. in orbiting, the ``meandering pattern''
determined by $\Omg/\omega$ will change depending on the
inhomogeneity strength.

In heart muscle, pinning of re-entrant waves of excitation has been
identified as a mechanism of conversion of ventricular fibrillation
into ventricular tachycardia~\cite{Valderrabano-etal-2000} and shown
to interfere with antitachycardia pacing~\cite{Ripplinger-etal-2006}.
Orbital movement is fundamentally 
more general than classical pinning:
in pinning, the spiral is attracted at all 
distances to the inhomogeneity of a certain sign, while
orbital movement occurs when repulsion at small distances changes
to attraction at larger distances, and this can happen at various 
distances and at either sign of inhomogeneity. The important practical consequense of the orbital movement is that a spiral may be bound
to inhomogeneities of either sign, even if it is repelled from
the inhomogeneity at small distances.

VNB is grateful to V.I.~Krinsky for inspiring discussions regarding
the problem of pinning.  This study has been supported in part by
EPSRC grants EP/D074789/1 and EP/D074746/1.


\begin{thebibliography}{28}
\expandafter\ifx\csname natexlab\endcsname\relax\def\natexlab#1{#1}\fi
\expandafter\ifx\csname bibnamefont\endcsname\relax
  \def\bibnamefont#1{#1}\fi
\expandafter\ifx\csname bibfnamefont\endcsname\relax
  \def\bibfnamefont#1{#1}\fi
\expandafter\ifx\csname citenamefont\endcsname\relax
  \def\citenamefont#1{#1}\fi
\expandafter\ifx\csname url\endcsname\relax
  \def\url#1{\texttt{#1}}\fi
\expandafter\ifx\csname urlprefix\endcsname\relax\def\urlprefix{URL }\fi
\providecommand{\bibinfo}[2]{#2}
\providecommand{\eprint}[2][]{\url{#2}}

\bibitem[{\citenamefont{Shagalov}(1997)}]{Shagalov-1997}
\bibinfo{author}{\bibfnamefont{A.~G.} \bibnamefont{Shagalov}},
  \bibinfo{journal}{Phys. Lett. A} \textbf{\bibinfo{volume}{235}},
  \bibinfo{pages}{643} (\bibinfo{year}{1997}).

\bibitem[{\citenamefont{Oswald and Dequidt}(2008)}]{Oswald-Dequidt-2008}
\bibinfo{author}{\bibfnamefont{P.}~\bibnamefont{Oswald}} \bibnamefont{and}
  \bibinfo{author}{\bibfnamefont{A.}~\bibnamefont{Dequidt}},
  \bibinfo{journal}{Phys. Rev. E} \textbf{\bibinfo{volume}{77}},
  \bibinfo{pages}{051706} (\bibinfo{year}{2008}).

\bibitem[{\citenamefont{Le~Berre et~al.}(2005)\citenamefont{Le~Berre, Ressayre,
  Tallet, and Tlidi}}]{LeBerre-etal-2005}
\bibinfo{author}{\bibfnamefont{M.}~\bibnamefont{Le~Berre}},
  \bibinfo{author}{\bibfnamefont{E.}~\bibnamefont{Ressayre}},
  \bibinfo{author}{\bibfnamefont{A.}~\bibnamefont{Tallet}}, \bibnamefont{and}
  \bibinfo{author}{\bibfnamefont{M.}~\bibnamefont{Tlidi}},
  \bibinfo{journal}{Phys. Rev. E} \textbf{\bibinfo{volume}{71}},
  \bibinfo{pages}{036224} (\bibinfo{year}{2005}).

\bibitem[{\citenamefont{Agladze and Steinbock}(2000)}]{Agladze-Steinbock-2000}
\bibinfo{author}{\bibfnamefont{K.}~\bibnamefont{Agladze}} \bibnamefont{and}
  \bibinfo{author}{\bibfnamefont{O.}~\bibnamefont{Steinbock}},
  \bibinfo{journal}{J.Phys.Chem. A} \textbf{\bibinfo{volume}{104 (44)}},
  \bibinfo{pages}{9816} (\bibinfo{year}{2000}).

\bibitem[{\citenamefont{Igoshin et~al.}(2004)\citenamefont{Igoshin, Welch,
  Kaiser, and Oster}}]{Igoshin-etal-2004}
\bibinfo{author}{\bibfnamefont{O.~A.} \bibnamefont{Igoshin}},
  \bibinfo{author}{\bibfnamefont{R.}~\bibnamefont{Welch}},
  \bibinfo{author}{\bibfnamefont{D.}~\bibnamefont{Kaiser}}, \bibnamefont{and}
  \bibinfo{author}{\bibfnamefont{G.}~\bibnamefont{Oster}},
  \bibinfo{journal}{Proc. Nat. Acad. Sci. USA} \textbf{\bibinfo{volume}{101}},
  \bibinfo{pages}{4256} (\bibinfo{year}{2004}).

\bibitem[{\citenamefont{Dahlem and M{\"u}ller}(2003)}]{Dahlem-Mueller-2003}
\bibinfo{author}{\bibfnamefont{M.~A.} \bibnamefont{Dahlem}} \bibnamefont{and}
  \bibinfo{author}{\bibfnamefont{S.~C.} \bibnamefont{M{\"u}ller}},
  \bibinfo{journal}{Biological Cybernetics} \textbf{\bibinfo{volume}{88}},
  \bibinfo{pages}{419} (\bibinfo{year}{2003}).

\bibitem[{\citenamefont{Bretschneider et~al.}(2009)}]{Bretschneider-etal-2009}
\bibinfo{author}{\bibfnamefont{T.}~\bibnamefont{Bretschneider}}
  \bibnamefont{et~al.}, \bibinfo{journal}{Biophys. J.}
  \textbf{\bibinfo{volume}{96}}, \bibinfo{pages}{2888} (\bibinfo{year}{2009}).

\bibitem[{\citenamefont{Fast and Pertsov}(1992)}]{Fast-Pertsov-1992}
\bibinfo{author}{\bibfnamefont{V.~G.} \bibnamefont{Fast}} \bibnamefont{and}
  \bibinfo{author}{\bibfnamefont{A.~M.} \bibnamefont{Pertsov}},
  \bibinfo{journal}{J. Cardiovasc. Electrophysiol.}
  \textbf{\bibinfo{volume}{3}}, \bibinfo{pages}{255} (\bibinfo{year}{1992}).

\bibitem[{\citenamefont{Luengviriya et~al.}(2006)\citenamefont{Luengviriya,
  Storb, Hauser, and M{\"u}ller}}]{Luengviriya-etal-2006}
\bibinfo{author}{\bibfnamefont{C.}~\bibnamefont{Luengviriya}},
  \bibinfo{author}{\bibfnamefont{U.}~\bibnamefont{Storb}},
  \bibinfo{author}{\bibfnamefont{M.~J.~B.} \bibnamefont{Hauser}},
  \bibnamefont{and} \bibinfo{author}{\bibfnamefont{S.~C.}
  \bibnamefont{M{\"u}ller}}, \bibinfo{journal}{Phys. Chem. Chem. Phys.}
  \textbf{\bibinfo{volume}{8}}, \bibinfo{pages}{1425} (\bibinfo{year}{2006}).

\bibitem[{\citenamefont{Nettesheim et~al.}(1993)\citenamefont{Nettesheim, von
  Oertzen, Rotermund, and Ertl}}]{Nettesheim-etal-1993}
\bibinfo{author}{\bibfnamefont{S.}~\bibnamefont{Nettesheim}},
  \bibinfo{author}{\bibfnamefont{A.}~\bibnamefont{von Oertzen}},
  \bibinfo{author}{\bibfnamefont{H.~H.} \bibnamefont{Rotermund}},
  \bibnamefont{and} \bibinfo{author}{\bibfnamefont{G.}~\bibnamefont{Ertl}},
  \bibinfo{journal}{J. Chem. Phys.} \textbf{\bibinfo{volume}{98}},
  \bibinfo{pages}{9977} (\bibinfo{year}{1993}).

\bibitem[{\citenamefont{Pertsov et~al.}(1993)\citenamefont{Pertsov, Davidenko,
  Salomonsz, Baxter, and Jalife}}]{Pertsov-etal-1993}
\bibinfo{author}{\bibfnamefont{A.~M.} \bibnamefont{Pertsov}},
  \bibinfo{author}{\bibfnamefont{J.~M.} \bibnamefont{Davidenko}},
  \bibinfo{author}{\bibfnamefont{R.}~\bibnamefont{Salomonsz}},
  \bibinfo{author}{\bibfnamefont{W.~T.} \bibnamefont{Baxter}},
  \bibnamefont{and} \bibinfo{author}{\bibfnamefont{J.}~\bibnamefont{Jalife}},
  \bibinfo{journal}{Circ. Res.} \textbf{\bibinfo{volume}{72}},
  \bibinfo{pages}{631} (\bibinfo{year}{1993}).

\bibitem[{\citenamefont{Lim et~al.}(2006)\citenamefont{Lim, Maskara, Aguel,
  Emokpae, and Tung}}]{Lim-etal-2006}
\bibinfo{author}{\bibfnamefont{Z.~Y.} \bibnamefont{Lim}},
  \bibinfo{author}{\bibfnamefont{B.}~\bibnamefont{Maskara}},
  \bibinfo{author}{\bibfnamefont{F.}~\bibnamefont{Aguel}},
  \bibinfo{author}{\bibfnamefont{R.}~\bibnamefont{Emokpae}}, \bibnamefont{and}
  \bibinfo{author}{\bibfnamefont{L.}~\bibnamefont{Tung}},
  \bibinfo{journal}{Circulation} \textbf{\bibinfo{volume}{114}},
  \bibinfo{pages}{2113} (\bibinfo{year}{2006}).

\bibitem[{\citenamefont{Lugomer et~al.}(2007)\citenamefont{Lugomer, Fukumoto,
  Farkas, Sz{\"or\'e}nyi, and Toth}}]{Lugomer-etal-2007}
\bibinfo{author}{\bibfnamefont{S.}~\bibnamefont{Lugomer}},
  \bibinfo{author}{\bibfnamefont{Y.}~\bibnamefont{Fukumoto}},
  \bibinfo{author}{\bibfnamefont{B.}~\bibnamefont{Farkas}},
  \bibinfo{author}{\bibfnamefont{T.}~\bibnamefont{Sz{\"or\'e}nyi}},
  \bibnamefont{and} \bibinfo{author}{\bibfnamefont{A.}~\bibnamefont{Toth}},
  \bibinfo{journal}{Phys. Rev. E} \textbf{\bibinfo{volume}{76}},
  \bibinfo{pages}{016305} (\bibinfo{year}{2007}).

\bibitem[{\citenamefont{Biktashev and Holden}(1995)}]{Biktashev-Holden-1995}
\bibinfo{author}{\bibfnamefont{V.~N.} \bibnamefont{Biktashev}}
  \bibnamefont{and} \bibinfo{author}{\bibfnamefont{A.~V.}
  \bibnamefont{Holden}}, \bibinfo{journal}{Chaos Solitons \& Fractals}
  \textbf{\bibinfo{volume}{5}}, \bibinfo{pages}{575} (\bibinfo{year}{1995}).

\bibitem[{\citenamefont{Biktasheva and
  Biktashev}(2003)}]{Biktasheva-Biktashev-2003}
\bibinfo{author}{\bibfnamefont{I.~V.} \bibnamefont{Biktasheva}}
  \bibnamefont{and} \bibinfo{author}{\bibfnamefont{V.~N.}
  \bibnamefont{Biktashev}}, \bibinfo{journal}{Phys. Rev. E}
  \textbf{\bibinfo{volume}{67}}, \bibinfo{pages}{026221}
  (\bibinfo{year}{2003}).

\bibitem[{\citenamefont{Biktasheva et~al.}(to be
  published)\citenamefont{Biktasheva, Barkley, Biktashev, and
  Foulkes}}]{Biktasheva-etal-2010}
\bibinfo{author}{\bibfnamefont{I.~V.} \bibnamefont{Biktasheva}},
  \bibinfo{author}{\bibfnamefont{D.}~\bibnamefont{Barkley}},
  \bibinfo{author}{\bibfnamefont{V.~N.} \bibnamefont{Biktashev}},
  \bibnamefont{and} \bibinfo{author}{\bibfnamefont{A.~J.}
  \bibnamefont{Foulkes}}, \emph{\bibinfo{title}{Computation of the drift
  velocity of spiral waves using response functions}} (\bibinfo{year}{to be
  published}).

\bibitem[{\citenamefont{Biktasheva et~al.}(2009)\citenamefont{Biktasheva,
  Barkley, Biktashev, Bordyugov, and Foulkes}}]{Biktasheva-etal-2009}
\bibinfo{author}{\bibfnamefont{I.~V.} \bibnamefont{Biktasheva}},
  \bibinfo{author}{\bibfnamefont{D.}~\bibnamefont{Barkley}},
  \bibinfo{author}{\bibfnamefont{V.~N.} \bibnamefont{Biktashev}},
  \bibinfo{author}{\bibfnamefont{G.~V.} \bibnamefont{Bordyugov}},
  \bibnamefont{and} \bibinfo{author}{\bibfnamefont{A.~J.}
  \bibnamefont{Foulkes}}, \bibinfo{journal}{Phys. Rev. E}
  \textbf{\bibinfo{volume}{79}}, \bibinfo{pages}{056702}
  (\bibinfo{year}{2009}).

\bibitem[{epa()}]{epaps}
\bibinfo{howpublished}{See EPAPS Document No. [number will be inserted by
  publisher] for a movie and details of numerical procedures. For more
  information on EPAPS, see {http://www.aip.org/pubservs/epaps.html}.}

\bibitem[{\citenamefont{Barkley}(1991)}]{Barkley-1991}
\bibinfo{author}{\bibfnamefont{D.}~\bibnamefont{Barkley}},
  \bibinfo{journal}{Physica D} \textbf{\bibinfo{volume}{49}},
  \bibinfo{pages}{61} (\bibinfo{year}{1991}).

\bibitem[{\citenamefont{Kness et~al.}(1992)\citenamefont{Kness, Tuckerman, and
  Barkley}}]{Kness-etal-1992}
\bibinfo{author}{\bibfnamefont{M.}~\bibnamefont{Kness}},
  \bibinfo{author}{\bibfnamefont{L.~S.} \bibnamefont{Tuckerman}},
  \bibnamefont{and} \bibinfo{author}{\bibfnamefont{D.}~\bibnamefont{Barkley}},
  \bibinfo{journal}{Phys. Rev. A} \textbf{\bibinfo{volume}{46}},
  \bibinfo{pages}{5054} (\bibinfo{year}{1992}).

\bibitem[{\citenamefont{Barkley}(1994)}]{Barkley-1994}
\bibinfo{author}{\bibfnamefont{D.}~\bibnamefont{Barkley}},
  \bibinfo{journal}{Phys. Rev. Lett.} \textbf{\bibinfo{volume}{72}},
  \bibinfo{pages}{164} (\bibinfo{year}{1994}).

\bibitem[{\citenamefont{LeBlanc and Wulff}(2000)}]{LeBlanc-Wulff-2000}
\bibinfo{author}{\bibfnamefont{V.~G.} \bibnamefont{LeBlanc}} \bibnamefont{and}
  \bibinfo{author}{\bibfnamefont{C.}~\bibnamefont{Wulff}}, \bibinfo{journal}{J.
  Nonlinear Sci.} \textbf{\bibinfo{volume}{10}}, \bibinfo{pages}{569}
  (\bibinfo{year}{2000}).

\bibitem[{\citenamefont{Elkin and Biktashev}(1999)}]{Elkin-Biktashev-1999}
\bibinfo{author}{\bibfnamefont{Y.~E.} \bibnamefont{Elkin}} \bibnamefont{and}
  \bibinfo{author}{\bibfnamefont{V.~N.} \bibnamefont{Biktashev}},
  \bibinfo{journal}{J. Biol. Phys} \textbf{\bibinfo{volume}{25}},
  \bibinfo{pages}{129} (\bibinfo{year}{1999}).

\bibitem[{\citenamefont{Paz{\'o} et~al.}(2004)\citenamefont{Paz{\'o}, Kramer,
  Pumir, Kanani, Efimov, and Krinsky}}]{Pazo-etal-2004}
\bibinfo{author}{\bibfnamefont{D.}~\bibnamefont{Paz{\'o}}},
  \bibinfo{author}{\bibfnamefont{L.}~\bibnamefont{Kramer}},
  \bibinfo{author}{\bibfnamefont{A.}~\bibnamefont{Pumir}},
  \bibinfo{author}{\bibfnamefont{S.}~\bibnamefont{Kanani}},
  \bibinfo{author}{\bibfnamefont{I.}~\bibnamefont{Efimov}}, \bibnamefont{and}
  \bibinfo{author}{\bibfnamefont{V.}~\bibnamefont{Krinsky}},
  \bibinfo{journal}{Phys. Rev. Lett.} \textbf{\bibinfo{volume}{93}},
  \bibinfo{pages}{168303} (\bibinfo{year}{2004}).

\bibitem[{\citenamefont{Valderrabano et~al.}(2000)\citenamefont{Valderrabano,
  Kim, Yashima, Wu, Karagueuzian, and Chen}}]{Valderrabano-etal-2000}
\bibinfo{author}{\bibfnamefont{M.}~\bibnamefont{Valderrabano}},
  \bibinfo{author}{\bibfnamefont{Y.~H.} \bibnamefont{Kim}},
  \bibinfo{author}{\bibfnamefont{M.}~\bibnamefont{Yashima}},
  \bibinfo{author}{\bibfnamefont{T.~J.} \bibnamefont{Wu}},
  \bibinfo{author}{\bibfnamefont{H.~S.} \bibnamefont{Karagueuzian}},
  \bibnamefont{and} \bibinfo{author}{\bibfnamefont{P.~S.} \bibnamefont{Chen}},
  \bibinfo{journal}{J. Amer. College of Cardiol.}
  \textbf{\bibinfo{volume}{36}}, \bibinfo{pages}{2000} (\bibinfo{year}{2000}).

\bibitem[{\citenamefont{Ripplinger et~al.}(2006)\citenamefont{Ripplinger,
  Krinsky, Nikolski, and Efimov}}]{Ripplinger-etal-2006}
\bibinfo{author}{\bibfnamefont{C.~M.} \bibnamefont{Ripplinger}},
  \bibinfo{author}{\bibfnamefont{V.~I.} \bibnamefont{Krinsky}},
  \bibinfo{author}{\bibfnamefont{V.~P.} \bibnamefont{Nikolski}},
  \bibnamefont{and} \bibinfo{author}{\bibfnamefont{I.~R.}
  \bibnamefont{Efimov}}, \bibinfo{journal}{Am. J. Physiol. --- Heart and Circ.
  Physiol.} \textbf{\bibinfo{volume}{291}}, \bibinfo{pages}{H184}
  (\bibinfo{year}{2006}).

\bibitem[{\citenamefont{Jensen et~al.}(1994)\citenamefont{Jensen, Pannbacker,
  Mosekilde, Dewel, and Borckmans}}]{Jensen-etal-1994}
\bibinfo{author}{\bibfnamefont{O.}~\bibnamefont{Jensen}},
  \bibinfo{author}{\bibfnamefont{V.~O.} \bibnamefont{Pannbacker}},
  \bibinfo{author}{\bibfnamefont{E.}~\bibnamefont{Mosekilde}},
  \bibinfo{author}{\bibfnamefont{G.}~\bibnamefont{Dewel}}, \bibnamefont{and}
  \bibinfo{author}{\bibfnamefont{P.}~\bibnamefont{Borckmans}},
  \bibinfo{journal}{Phys. Rev. E} \textbf{\bibinfo{volume}{50}},
  \bibinfo{pages}{736} (\bibinfo{year}{1994}).

\bibitem[{\citenamefont{Zou et~al.}(1993)\citenamefont{Zou, Levine, and
  Kessler}}]{Zou-etal-1993}
\bibinfo{author}{\bibfnamefont{X.}~\bibnamefont{Zou}},
  \bibinfo{author}{\bibfnamefont{H.}~\bibnamefont{Levine}}, \bibnamefont{and}
  \bibinfo{author}{\bibfnamefont{D.~A.} \bibnamefont{Kessler}},
  \bibinfo{journal}{Phys. Rev. E} \textbf{\bibinfo{volume}{47}},
  \bibinfo{pages}{R800} (\bibinfo{year}{1993}).

\end{thebibliography}

\newpage

\makeatletter
  \onecolumngrid {\centering
  {\large\bfseries
    Orbital movement of spiral waves \\
    Appendix: Details of numerical methods and procedures \\[\baselineskip]
  }
  V.~N.~Biktashev \\
  {\it Department of Mathematical Sciences, University of Liverpool, Liverpool L69 7ZL, UK} \\[\baselineskip]
  D.~Barkley \\
  {\it Mathematics Institute, University of Warwick, Coventry CV4 7AL, UK} \\[\baselineskip]
  I.~V.~Biktasheva \\
  {\it Department of Computer Science, University of Liverpool, Liverpool L69 3BX, UK} \\[\baselineskip]
  } \twocolumngrid

\paragraph{Calculations of the spiral wave solution $\U$ and the response function
$\RF$.} 
We used a polar grid in a disk of radius $12.8$,  discretized
with 320 intervals in the radial direction and 128 intervals in the
angular direction.

\paragraph{Direct numerical simulations.}

We used forward Euler timestepping with $\dt=0.00128$, five-point
approximation of the Laplacian with $\dx=0.08$  and no-flux
  boundary conditions in a rectangular domain. The size of the domain
  was chosen big enough so that further increase did not change the
  behaviour. Typical sizes were $24\times24$ and $28\times24$. Typical
  initial conditions were $u(x,y,0)=\Heav(x-\xs)$,
  $v(x,y,0)=-b+a\Heav(y-\ys)$ where $\xs,\ys$ were chosen depending on
  the desired position of the spiral wave. 

The tip was defined as the point where $u=0.5$ and $v=0.25$ at the
given time, using bilinear interpolation between the grid nodes.  The
angle of $\nabla u$ at the tip with respect to $x$ axis,
calculated using the same interpolation, was taken as its
orientation. Positions of the centers were calculated by averaging the
tip position during the time intervals when the orientation made the full
circle $(-\pi,\pi]$.

The angular velocity of the orbital drift $\Omg$ was calculated as
$2\pi/\P$, where $\P$ was the simulation time taken for one whole turn
of the orbital drift, calculated to the nearest spiral rotation
period.

\end{document}